\documentclass[reprint,superscriptaddress,twocolumns,notitlepage,floatfix]{revtex4-1}
\usepackage[utf8]{inputenc}
\usepackage[T1]{fontenc}

\usepackage{float}
\usepackage{physics}
\usepackage{graphicx}
\usepackage{dcolumn}
\usepackage{bm}
\usepackage[x11names]{xcolor}
\usepackage{booktabs}
\usepackage{dsfont}
\usepackage{mathrsfs}
\usepackage{amssymb}
\usepackage[normalem]{ulem}
\usepackage{tabularx}

\usepackage{xr}
\newcommand{\shortspace}{\enspace\enspace}
\newcommand{\midspace}{\shortspace\shortspace}
\newcommand{\longspace}{\midspace\midspace}

\newcommand{\atom}{\text{a}}
\newcommand{\func}{\text{f}}
\newcommand{\mol}{\text{m}}
\newcommand{\diff}{\text{d}}

\newcommand{\cc}{\mathbf{c}}
\newcommand{\h}{\mathbf{h}}
\newcommand{\x}{\mathbf{x}}
\newcommand{\y}{\mathbf{y}}

\newcommand{\key}{\mathbf{k}}
\newcommand{\que}{\mathbf{q}}

\newcommand{\MLP}{\text{MLP}}
\newcommand{\MPN}{\text{MPN}}
\newcommand{\softmax}{\text{softmax}}

\newcommand{\G}{\mathcal{G}}
\newcommand{\V}{\mathcal{V}}
\newcommand{\E}{\mathcal{E}}

\newcommand{\Ga}{\G^\atom}
\newcommand{\Va}{\V^\atom}
\newcommand{\Ea}{\E^\atom}
\newcommand{\ha}{\h^\atom}
\newcommand{\xa}{\x^\atom}

\newcommand{\Gf}{\G^\func}
\newcommand{\Vf}{\V^\func}
\newcommand{\Ef}{\E^\func}
\newcommand{\hf}{\h^\func}
\newcommand{\xf}{\x^\func}

\newcommand{\hmol}{\h^\mol}

\begin{document}

\title{Attention-Based Functional-Group Coarse-Graining: \\
A Deep Learning Framework for Molecular Prediction and Design}

\author{Ming Han}
\thanks{These authors contributed equally to this work.}
\affiliation{Pritzker School of Molecular Engineering,
  University of Chicago, Chicago, IL, USA}
\affiliation{James Franck Institute,
  University of Chicago, Chicago, IL, USA}

\author{Ge Sun}
\thanks{These authors contributed equally to this work.}
\affiliation{Pritzker School of Molecular Engineering,
  University of Chicago, Chicago, IL, USA}
\affiliation{Department of Chemical and Biomolecular Engineering, Tandon School of Engineering, New York University, Brooklyn, NY, USA}
\affiliation{Department of Computer Science, Courant Institute of Mathematical Sciences, New York University, New York, NY, USA}
\affiliation{Department of Physics, New York University, New York, NY, USA}

\author{Juan J. de Pablo}
\thanks{Corresponding author. Email: jjd8110@nyu.edu}
\affiliation{Pritzker School of Molecular Engineering,
  University of Chicago, Chicago, IL, USA}
\affiliation{Department of Chemical and Biomolecular Engineering, Tandon School of Engineering, New York University, Brooklyn, NY, USA}
\affiliation{Department of Computer Science, Courant Institute of Mathematical Sciences, New York University, New York, NY, USA}
\affiliation{Department of Physics, New York University, New York, NY, USA}
\affiliation{Materials Science Division, Argonne National Laboratory, Lemont, IL, USA}

\begin{abstract}
\textbf{Abstract}: 
Machine learning (ML) offers considerable promise for the design of new molecules and materials. In real-world applications, the design problem is often domain-specific, and suffers from insufficient data, particularly labeled data, for ML training. In this study, we report a data-efficient, deep-learning framework for molecular discovery that integrates a coarse-grained functional-group representation with a self-attention mechanism to capture intricate chemical interactions. Our approach exploits group-contribution theory to create a graph-based intermediate representation of molecules, serving as a low-dimensional embedding that substantially reduces the data demands typically required for training. By leveraging the self-attention mechanism to learn subtle chemical context, our method consistently outperforms conventional methods in predicting multiple thermophysical properties. In a case study focused on adhesive polymer monomers, we train on a limited dataset comprising just 6,000 unlabeled and 600 labeled monomers. The resulting chemistry prediction model achieves over 92\% accuracy in forecasting properties directly from SMILES strings, exceeding the performance of current state-of-the-art techniques. Furthermore, the latent molecular embedding is invertible, allowing the design pipeline to incorporate a decoder that can automatically generate new monomers from the learned chemical subspace. We illustrate this functionality by targeting high and low glass transition temperatures (\(T_g\)), successfully identifying novel candidates whose \(T_g\) extends beyond the range observed in the training data. The ease with which our coarse-grained, attention-based framework navigates both chemical diversity and data scarcity offers a compelling route to accelerate and broaden the search for functional materials.

\end{abstract}
\maketitle

Molecular design lies at the core of modern science and engineering~\cite{sanchez2018inverse}, with wide applications that range from development of new drugs~\cite{patani1996bioisosterism,anderson2003process,vamathevan2019applications, stokes2020deep, nigam2023tartarus} to discovery of new functional and sustainable materials~\cite{beaujuge2011molecular,butler2018machine,merchant2023scaling,yao2023machine,gurnani2024ai}. 
While considerable progress has been made over decades of sustained efforts, it continues to be a daunting endeavor. Constructing a molecule with specific target properties involves a combinatorial problem that consists of selecting the correct atoms and connecting them in an appropriate manner.
The available chemical space for molecule design grows exponentially with molecular size~\cite{hansen2015machine}. Relevant candidates, however, only populate a very small portion of that space.
To identify optimal choices, it is necessary to address two crucial questions: the relationship between different molecular structures and the dependence of molecular properties on them. 
This gives rise to the inherent challenge of exploring chemical space, further exacerbating the \textit{curse of dimensionality} that pervades molecular design.

Molecular embedding~\cite{cereto2015molecular,duvenaud2015convolutional,coley2017convolutional,tshitoyan2019unsupervised} can facilitate the navigation of chemical space.
By evaluating a selection of molecular features, we can encode a molecule $M$ by its corresponding feature vector $\h$, denoted as:
\begin{equation}
   \mathcal{R}: M \to \h.
\end{equation} 
This mapping creates a mathematical realization of the chemical space, where the differences between molecules can be quantified via the distance $\lVert\h_i - \h_j\rVert$, and molecular properties can be inferred from a function $\y = f(\h)$. 
A good molecular embedding should satisfy the two following requirements. First, it must be chemically meaningful. Molecules with similar chemistry should be arranged close to each other, so that one can explore primarily relevant regions.
Second, it should be informative. The feature vector should contain key information for the prediction of molecular properties, which in turn can provide guidance for the optimization of molecular structure.

Molecular fingerprints~\cite{morgan1965generation,glen2006circular} are often used in traditional cheminformatics. They are prescribed descriptors that record the statistics of the different chemical groups in a molecule. This type of embedding organizes molecules in chemical space based on their local structures, which play a key role in determining molecular properties.
With that, if a new molecule with a new set of properties was sought, one could sample new candidates from existing molecules by simply replacing chemical groups and then screening the proposed constructs by using chemistry prediction models, such as Quantitative Structure-Activity Relationships (QSARs)~\cite{nantasenamat2009practical,khan2016descriptors}.
Such an approach, however, is limited to the exploration of the space in the near vicinity of individual known molecules, which is what local modifications allow. 
Furthermore, the quality of the resulting designs can be compromised by artifacts resulting from the interplay between chemical groups, especially their inter-connectivity, which is disregarded by molecular fingerprints. 

Recent advances in machine learning (ML) have enabled the extraction of molecular embedding from data~\cite{elton2019deep,david2020molecular,walters2020applications,wigh2022review,li2022deep}. A widely adopted ML scheme relies on an autoencoder~\cite{kingma2013auto,lecun2015deep} architecture, where an encoder maps the molecules into a continuous latent space, and a consecutive decoder tries to reconstruct them~\cite{gomez2018automatic,jin2018junction}. The following schematic provides a simple description of the key concepts:

\begin{figure}[H]
\includegraphics[width=0.48\textwidth]{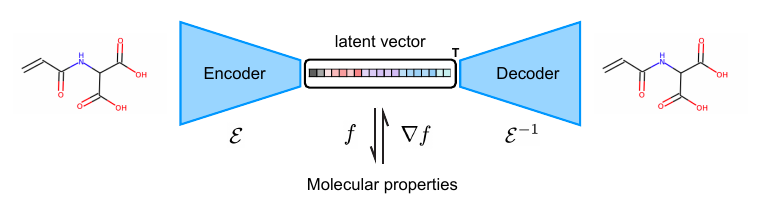}
\end{figure}

After being trained, an autoencoder can produce latent vectors that represent the global structure of a molecule.
This embedding enables coherent non-local modifications of molecules.
Thanks to the continuity of the latent space, interpolations can be made between different molecules to acquire combined properties, and directed optimization can be performed through gradient descent.
However, since this embedding is primarily designed for molecular reconstruction, it does not necessarily correlate well with molecular properties. Molecular reconstruction focuses on the connectivity of atoms or chemical groups within a molecule, whereas molecular chemistry is also influenced by the interactions between different molecules.
To enhance such correlations, the autoencoder should be jointly trained with an additional chemistry prediction model that maps latent vectors onto properties of interest.
Doing so would require a large amount of labeled data, which is typically unavailable. In real-world applications, the design problem is often domain-specific; there is a preferred class of molecules, and the subset for which properties have been measured or are available is very limited.

In this work, we introduce a machine-learning pipeline for domain-specific molecular design, anchored by a functional-group-based coarse-graining strategy. The pipeline consists of a hierarchical coarse-grained graph autoencoder to generate relevant candidates, and a chemistry prediction model to efficiently screen the proposed structures and select optimal choices. A key innovation here is the integration of a self-attention mechanism, inspired by natural language processing—where tokens in a sequence can have long-range dependencies—into the realm of macromolecules, whose functional groups exhibit similarly intricate spatial and chemical interactions. We anticipate broad applicability of this framework wherever faithful molecular generation is essential, from pharmaceutical discovery—where scaffold and functional-group placements critically affect bioactivity—to materials science, where chain-like and branched architectures frequently govern mechanical and thermal properties. Crucially, by focusing on functional-group-based coarse-graining, our hierarchical approach remains data-efficient, enabling robust design and analysis even under data-scarce conditions.

\vspace{5mm}

\textbf{Coarse-grained graph autoencoder}
\smallskip

Depending on the choice of molecular representation, there are two popular ways to create an autoencoder. 
By representing a molecule as a SMILES~\cite{weininger1988smiles} string, we can treat molecular embedding as a natural language processing problem and build an autoencoder using a sequence model~\cite{gomez2018automatic}. 
But due to the one-dimensional nature of a text string, special tricks are always needed to account for the three-dimensional topology of a molecule, especially for features like rings, branches, and stereoisomerism. 
This poses unnecessary obstacles to the application of string-based autoencoders. 
Alternatively, by representing a molecule as a graph of atoms, we can naturally preserve molecular topology and embed molecules using graph neural networks. 
In fact, atom-graph-based autoencoders have been widely used for the design of small-molecule drugs. 
However, since these autoencoders generate molecules atom-by-atom, they suffer from information loss when scaled up for the description of large molecules.

\begin{figure*}
\includegraphics[width=0.98\textwidth]{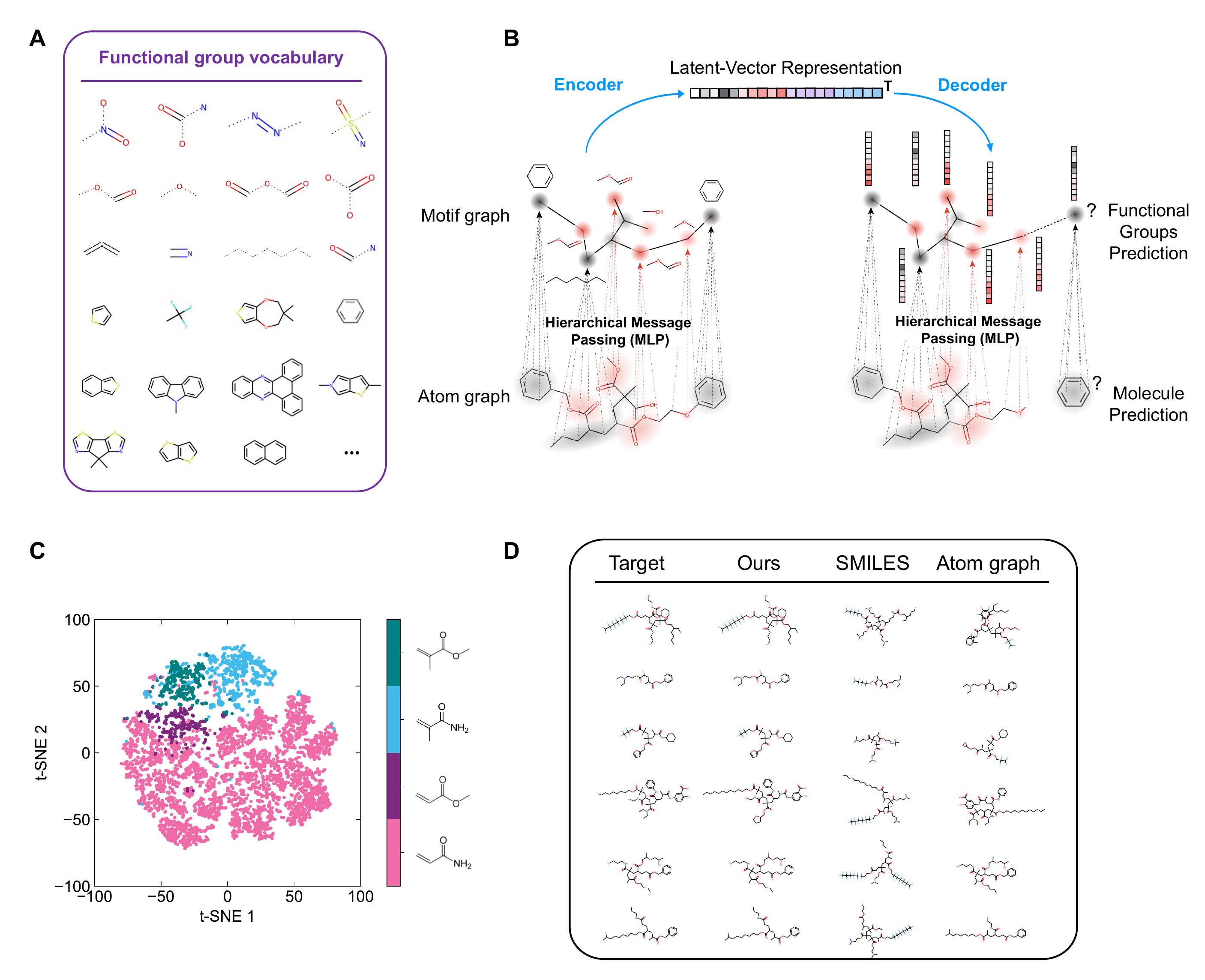}\caption{\textbf{Coarse-grained autoencoder using hierarchical graph neural networks.} 
By identifying "elementary" functional groups, we can construct a hierarchical representation of a molecule, with an atom graph at the finer level and a motif graph at the coarser level.
\textbf{(A)}  Vocabulary of functional groups. The vocabulary is composed of 50 key chemical groups selected on the basis of group contribution theory, as well as all the ring structures that appear in the known molecules of interest. \textbf{(B)} Coarse-grained graph autoencoder. Graph neural networks are applied to both the atom and motif graphs to encode the local environment of individual nodes, which contains the nodes themselves and their neighbors. A multilayer perceptron (MLP) is introduced between two graphs to integrate the encoded information at the atom level into functional groups at the motif level. A variational connection is added to ensure the continuity of the latent space. Note that the decoder operates in an autoregressive manner. It reconstructs the molecule by iteratively generating a new motif and connecting it to the partial molecule built thus far. At each step, it needs an encoder to refresh the embedding of atoms and functional groups based on the current molecular structure. \textbf{(C)} Chemically meaningful embeddings from our autoencoder. This visualization demonstrates the capability of our autoencoder to categorize 6000 adhesive monomers into four clusters corresponding to their chemical classes: Methyl Methacrylate, Methacrylamide, Methyl Acrylate, and Acrylamide. For illustrative purposes, we use t-SNE to map ten-dimensional embeddings into a two-dimensional space.  Even though the types of monomers were not provided during training, our autoencoder can automatically create embeddings that show clear separations based on inherent chemical properties. \textbf{(D)} Our embeddings are also invertible. We compare our functional-group-based autoencoder with others built on SMILES and atom graph representations. Molecules used in industrial materials applications often contain more chains and fewer rings than small-molecule drugs. For these molecules, our autoencoder can still deliver 95\% reconstruction accuracy, whereas those based on SMILES and atom graphs only offer a 60\% or lesser accuracy. 
} \label{fig:autoencoder}
\end{figure*}

Constructing molecules using structural motifs provides an effective means for the design of large molecules having a complex topology. Researchers have identified approximately 100 functional groups, which are local structures that underlie the key chemical properties of molecules. Notably, most synthesizable molecules can be deconstructed into these structural motifs. Thus, this small set of common functional groups (as shown in Fig.~\ref{fig:autoencoder}A) can serve as a standard vocabulary for molecular design. Compared to atoms, they enable a coarse-grained and chemically meaningful representation of a molecule, which simplifies the design process.

Expanding upon recent advances in hierarchical encoder-decoder~\cite{jin2020hierarchical} architectures for molecular graphs, we constructed our functional-group-based autoencoder around a multi-level representation of molecular structures. As illustrated in Fig.~\ref{fig:autoencoder}A, a molecule $M$ can be represented with two levels of description: at the fine level, it forms an atom graph $\Ga(M)$ composed of atoms $a_i$ and bonds $b_{ij}$; at the coarse level, it is also a motif graph $\Gf(M)$ composed of functional groups $F_u$ and their interconnectivity $E_{uv}$; in between is the hierarchical mapping from each functional group $F_u$ towards the corresponding atomic subgraph $\Ga(F_u)$, which is easily accessible via the cheminformatics software RDKit~\cite{rdkit}. 
Details are summarized below: (Throughout the paper, superscripts and subscripts are used to denote the hierarchical level and the node index within a graph, respectively.)

\onecolumngrid
\begin{table}[H]
\centering
\begin{tabularx}{\textwidth} {>{\centering}X >{\centering\arraybackslash}X >{\centering\arraybackslash}X >{\centering\arraybackslash}X}
    \toprule
    \multicolumn{4}{c}{\bf Coarse-graining graph representation}\\
    \midrule
    $\begin{aligned}
        &\;\;\textbf{Motif level} \\[0.4mm]
        &\textbf{Motif} \to {\bf Atom} \\[0.4mm]
        &\;\;\,\textbf{Atom level}
    \end{aligned}$
    &
    $\begin{aligned}
        \Gf(M) &= \big(\Vf(M), \Ef(M)\big) \\
        \Ga(F_u) &= \left(\Va(F_u), \Ea(F_u)\right) \\
        \Ga(M) &= \big(\Va(M), \Ea(M)\big)
    \end{aligned}$
    &
    $\begin{aligned}
        \Vf(M) &= \left\{F_u \,|\, F_u \in M \right\} \\[0.4mm]
        \Va(F_u) &= \left\{a_i \,|\,a_i \in F_u \right\} \\[0.4mm]
        \Va(M) &= \left\{a_i \,|\, a_i \in M \right\}
    \end{aligned}$
    &
    $\begin{aligned}
        \Ef(M) &= \left\{E_{uv} \,|\, E_{uv} \in M \right\} \\[0.4mm]
        \Ea(F_u) &= \left\{b_{ij} \,|\, b_{ij} \in F_u \right\} \\[0.4mm]
        \Ea(M) &= \left\{b_{ij} \,|\, b_{ij} \in M \right\}
    \end{aligned}$
    \\
    \bottomrule
\end{tabularx}
\end{table}
\twocolumngrid

Molecular embedding is introduced by treating the generation of molecules as a Bayesian inference:
\begin{equation}
P(M) = \int \diff \hmol \,P(\hmol)  \, P \big( M \, \bigr| \, \hmol \big) \label{eq:prob_mol}.
\end{equation}
$P(\hmol)$ is a prior distribution of the embedding $\hmol$. In the following text, we will explain in more detail how to build an encoder to estimate the posterior distribution $P(\hmol|M)$ and a decoder to derive the conditional probability of reconstructing the same molecule $P ( M | \hmol)$.

The encoder analyzes a molecule in a bottom-up fashion. First, a message-passing network ($\MPN$) is used to encode the atom graph:
\begin{align}
\left\{ \ha_i \right\}, \, \left\{ \ha_{ij} \right\} = \MPN^\atom \Big( \Ga, \, \left\{\xa_i\right\}, \,\left\{\xa_{ij}\right\} \Big). \label{eq:GNN_atom}
\end{align}
Here the features of individual atoms and bonds, for instance, $\xa_i=$ (atom type, valence, formal charge) and $\xa_{ij} =$ (bond type, stereo-chemistry) are taken as inputs and shared with their neighbors through the message passing mechanism on the graph $\Ga$. Then the embeddings of the atoms and bonds are derived to encode their local environment, denoted as $\ha_i$ and $\ha_{ij}$, which include their own properties and those of their neighbors, as well as the way in which they connect with each other. Second, we assemble the feature vectors of functional groups using a multi-layer perceptron (MLP): 
\begin{align} 
\xf_u = \MLP^{\atom \to \func} \Big(\mathbf{x}(F_u), \sum_{i \,\in \Va(F_u)} \ha_{i}\Big),
\end{align}
which consists of the type of embedding of the functional group $\mathbf{x}(F_u)$ and the graph embedding of its atomic components $\big\{ \ha_{i} \, \bigr|\, i \in \Va(F_u)\big\}$. Third, we employ another MPN to encode the motif graph:
\begin{align}
\left\{ \hf_u \right\}, \, \left\{ \hf_{uv} \right\} = \MPN^\func \Big( \Gf, \, \left\{\xf_u\right\}, \,\left\{\xf_{uv}\right\} \Big) 
\end{align}
where $\hf_{u}$ and $\hf_{uv}$ are the embeddings of individual functional groups. Lastly, a molecular embedding $\hmol$ is sampled in a probabilistic manner:
\begin{equation}
    P(\hmol|M) \approx \mathcal{N}\Big(\boldsymbol{\mu}(\hf_0), \,\boldsymbol{\sigma}(\hf_0)\Big)
    \label{eq:encoder}
\end{equation}
where both the mean $\boldsymbol{\mu}$ and standard deviation $\boldsymbol{\sigma}$ of the above normal distribution are derived from the embedding $\hf_0$ of a leaf node on graph $\Gf$. Besides generating an estimate on the posterior distribution $P(\hmol|M)$, this variational design also ensures the continuity of the latent space.

Once the molecular embedding is achieved, the decoder can reconstruct the same molecule motif-by-motif. We model this as an auto-regressive progress by factorizing the conditional probability in Eq.~\eqref{eq:prob_mol} into:
\begin{equation}
    P(M|\,\hmol) = \prod_{u} P \Big( M_{\leq u} \, \Bigr| \, M_{\leq u-1}, \, \hmol \Big)
    \label{eq:decoder}
\end{equation}
where $M_{\leq u-1} = \bigcup_{v \leq u-1}F_v$ and $M_{\leq u} = \bigcup_{v \leq u}F_v$ denote the molecule before and after adding functional group $F_u$, respectively. At each step, the decoder only needs to predict which functional group $F_u$ to choose and which bond $b_{ij}$ to form for its attachment onto the partial molecule $M_{\leq u-1}$:
\begin{equation}
\begin{aligned}
P \Big( M_{\leq u} \, \Bigr|\, *\Big) &= P\Big(b_{ij} \, \Bigr| \, F_u, \, *\, \Big) \, P \Big(F_u \, \Bigr| \, * \Big) \\[1mm]
\text{with} \midspace * \;\;\; &= \big(M_{\leq u-1}, \;\hmol\big). \label{eq:cond_prob}
\end{aligned}
\end{equation}

To quantify the condition, we apply the same encoder to analyze $M_{\leq u-1}$.  Assuming that the useful information on the partial molecule for the prediction of the next motif is localized near its growing end, we use the embedding of the last motif $\hf_{u-1}$ to represent $M_{\leq u-1}$. The probability of choosing functional group $F_u$ as the next motif can then be estimated as
\begin{equation}
\begin{aligned}
& P \Big(F_u \, \Bigr| \, M_{\leq u-1}, \hmol \Big) \, \approx \, P \Big(F_u \, \Bigr|\, \hf_{u-1}, \, \hmol \Big) \\[1mm]
& \midspace =  \softmax\Big( \, \MLP\,(\hf_{u-1}, \, \hmol) \,\Big).
\end{aligned}
\end{equation}
Similarly, for the prediction of the attachment bond,
atom embeddings $\big\{\ha_i\,\bigr|\,a_i \in F_u\big\}$ and $\big\{\ha_j\,\bigr|\,a_j \in F_{u-1}\big\}$ are used to represent the condition on both $F_u$ and $M_{\leq u-1}$. Then we estimate the probability of choosing bond $b_{ij}$ as
\begin{equation}
\begin{aligned}
& P \Big(b_{ij} \, \Bigr| \, F_u,  \, M_{\leq u-1}, \, \hmol \Big) \approx P \Big(b_{ij} \, \Bigr| \, \ha_i, \, \ha_j, \, \hmol \Big)  \\
&\longspace\; = \softmax\Big(\, \MLP\, (\ha_i, \, \ha_j, \, \hmol )\, \Big).
\label{eq:end_decoder}
\end{aligned}
\end{equation}
Since $F_u$ is not yet attached to $M_{\leq u-1}$, here the embedding $\ha_i$ is obtained by applying the encoder to $F_u$ alone. 

We train our model by minimizing its evidence lower bound (ELBO), a common loss function for a variational autoencoder. It contains two parts $\mathcal{L}_\text{ELBO} = \mathcal{L}_1 + \lambda \mathcal{L}_2$, namely 
a reconstruction loss that quantifies the cross-entropy between the encoder and the decoder 
\begin{equation}
\mathcal{L}_1 = H\Big[P(\hmol|M), P(M|\,\hmol)\Big]
\end{equation}
and a regularizer that measures the Kullback–Leibler (KL) divergence between the prior and posterior distributions of molecular embedding to avoid overfitting 
\begin{equation}
\mathcal{L}_2 = \mathcal{D}_\text{KL} \Big[P(\hmol|M) \,\bigr|\bigr|\,P(\hmol)\Big] \,
\end{equation}
where we postulate the prior distribution $P(\hmol) = \mathcal{N}(\mathbf{0}, \,\mathbf{I})$, and where the probability estimates by the encoder and the decoder, $P(\hmol|M)$ and $P(M|\,\hmol)$, are computed using Eqs.~(\ref{eq:encoder}-\ref{eq:decoder}).

To test the performance of the autoencoder, we apply it to embedding monomers for the design of acrylate-based adhesive polymeric materials. In this domain-specific application, 6000 known molecules of 4 different monomer types are chosen to form the dataset. As an unsupervised learning tool, the autoencoder does not require data to be labeled. Thus no molecular properties are provided here. Yet, our autoencoder automatically clusters the monomers based on their types, as shown by the t-SNE projection of the latent space $\mathcal{S}(\hmol)$, see Fig.~\ref{fig:autoencoder}C. This could be beneficial to the directed design of molecules with targeted chemical properties. 

Besides furnishing a chemically meaningful representation, our molecular embedding also remains fully invertible—a vital feature for generative tasks. Specifically, Using the embedding of a molecule produced by the encoder, the decoder can reconstruct the same molecule with an accuracy of approximately 95\%, substantially surpassing string-based and atom-graph-based autoencoders \cite{gomez2018automatic,liu2018constrained} trained under comparable conditions. Although previous studies \cite{jin2020hierarchical, jin2018junction} reported decent reconstruction rates on broader chemical libraries, those architectures were tailored to relatively diverse molecules in which rings are prevalent. Here, by contrast, we seek to capture the richer chain- and branch-dominated chemistry of polymeric adhesives, where ring motifs are relatively scarce. In such data-limited, domain-specific settings, we observed that traditional decomposition schemes—which often mine ring and branch substructures based purely on occurrence frequency—induce an unbalanced motif vocabulary heavily biased toward rings. This bias not only hampers the model’s ability to reconstruct chain-like polymers accurately but also leads to less chemically interpretable embeddings as shown in Supplementary Fig. S1. 

Our approach addresses this bottleneck by deliberately constructing a motif vocabulary from functional groups, building on the established principles of group contribution. This strategy provides two main advantages. First, it limits our dictionary to functional groups that comprehensively span typical polymeric architectures, mitigating overfitting to ring motifs. Second, by focusing on functional groups rather than purely structural motifs, the model inherently encodes information relevant to reactivity and physical properties, resulting in embeddings that more closely align with chemical intuition. The net effect is markedly superior reconstruction fidelity, particularly for the class of polymeric monomers under study. Such a design aligns well with the aims of data-scarce, domain-focused molecular design, wherein capturing specialized chemistries effectively can be more important than achieving broad coverage of all possible small-molecule scaffolds. Consequently, the strong reconstruction accuracy reflects our model’s capacity to handle polymer-like structures with minimal data, highlighting the promise of functional-group-based coarse-graining for improved generative performance.

\vspace{5mm}

\textbf{Attention-aided chemistry prediction model}
\smallskip

\begin{figure*}
\includegraphics[width=0.9\textwidth]{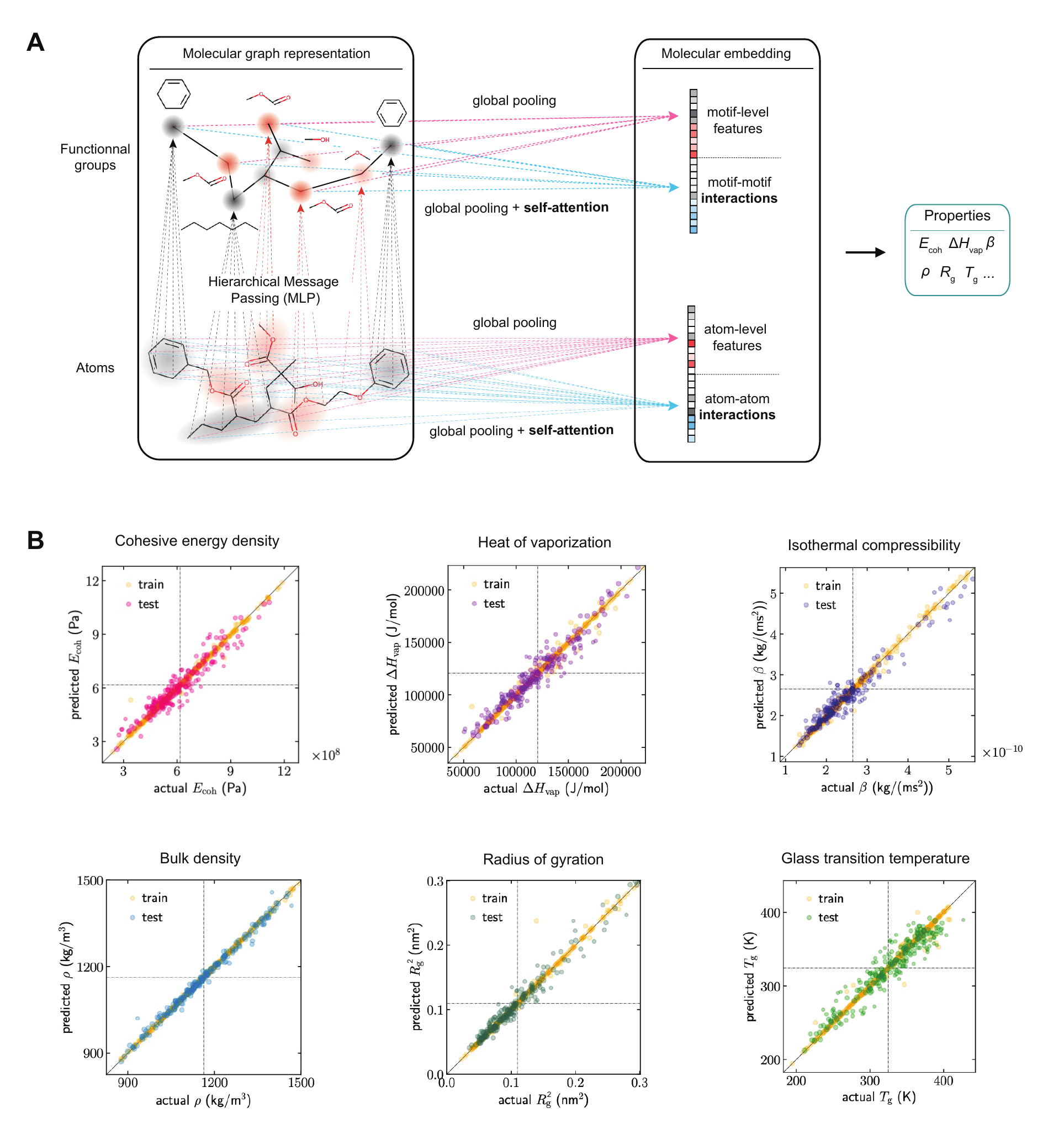} \caption{\textbf{Chemistry prediction model.} \textbf{(A)} Chemistry-oriented molecular embedding. Instead of using the latent vector for molecular reconstruction, we develop another molecular embedding focused on chemical properties. It has the same architecture as the encoder mentioned above. The key difference lies in the use of global pooling to reduce graph size and extract molecular embedding with fixed length, rather than using the embedding of the root motif. Two types of pooling are employed here. A direct graph pooling with sigmoid rectification is used to summarize the contributions of individual nodes. A pooling layer with a self-attention mechanism is also used to summarize the contributions of node-node interactions. The chemistry-oriented molecule embedding is the concatenation of individual and interaction embeddings at both the atom and motif levels. We can then use the final embedding to predict molecular properties of interest. \textbf{(B)} Prediction performance. We apply this model to learn the dependence of monomer properties on their molecular structures. Six properties relevant to the design of adhesive materials are considered and obtained using automated molecular dynamics simulations. The model is trained on 400 labeled monomers (yellow dots) and tested on another, unseen 200 monomers (other colored dots). Except for glass transition temperature, all other properties can be predicted with high accuracy $R^2 > 0.92$.} \label{fig:chem}
\end{figure*}

Our autoencoder imposes a relationship between different molecular structures, by mapping them into a continuous latent space where they are organized in a chemically meaningful manner. This allows for efficient sampling of chemically relevant candidates. To achieve directed design, an efficient approach is still needed to evaluate candidate properties. While high throughput experiments or simulations can consider hundreds of molecular species at a reasonable cost, they are limited in terms of scalability. To address this challenge, we propose a chemistry prediction model capable of directly deriving thermophysical properties from molecular structures by using the self-attention mechanism.

The model is built on the same coarse-grained graph representation of a molecule, shown in Fig.~\ref{fig:chem}A. Instead of being a simple readout from the molecular embedding, it analyzes the embedding of individual atoms and functional groups. Compared to the global structure of a molecule, the variation of those local structures is much more constrained. Therefore, training a regression model on the latter is more data-efficient. This is particularly useful for domain-specific design, where labeled data are limited. 

We first estimate the contributions of individual atoms and functional groups to molecular properties, by applying the following equation to the corresponding level of the graph hierarchy:
\begin{align}
\cc_i = \text{Concat}\left(\mathbf{\hat{c}}(\h_i), \; \sum_j a_{ij} \,\mathbf{\Tilde{c}} (\h_i, \h_j)\right), \label{eq:contribution}
\end{align}

where $\mathbf{\hat{c}}(\h_i)$ represents the contribution of node $i$ alone, while $\mathbf{\Tilde{c}}(\h_i, \h_j)$ represents the contribution of the interaction between node $i$ and node $j$ that can be weighted by a learnable coefficient, \( a_{ij} \). Nodes can refer to either atoms or functional groups depending on the level of the graph, with $\h_i = \ha_i \text{ or } \hf_i$. 

In bulk materials, the likelihood of two local structures interacting with each other is determined by their affinity. To capture this relationship, we introduce the weight coefficient, \( a_{ij} \), defined by using the multi-head attention mechanism~\cite{vaswani2017attention}:
\begin{align}
a_{ij} = \text{softmax}\left(\frac{\que_i\key_j^T}{\sqrt{d_k}}\right) = \frac{\exp\left(\frac{\que\left(\h_i\right) \cdot \key\left(\h_j\right)^T}{\sqrt{d_k}}\right)}{\sum_{l} \exp\left(\frac{\que\left(\h_i\right) \cdot \key\left(\h_l\right)^T}{\sqrt{d_k}}\right)}, \label{eq:attention}
\end{align}
The attention weights $a_{ij}$ are computed using the query, key, and value projections derived from the node representation \( \h_i \). In this context, the query $\que(\h_i)$ represents the features that node $i$ requests from its interaction counterparts, while the key $\key(\h_j)$ represents the features that node $j$ possesses and can match. The dimension of the keys is denoted by $d_k$. The affinity score between nodes $i$ and $j$ is obtained by performing a dot product of their features, which is then scaled by $1/\sqrt{d_k}$ to stabilize the gradients during training. This scaled dot-product is transformed into attention by applying the softmax function to normalize the influence of each interaction. Finally, the resulting attention weight $a_{ij}$ can be interpreted as the probability of node $i$ interacting with node $j$, taking into account all other nodes in the graph.

We can then derive the molecular properties from the contributions of individual atoms and functional groups, denoted as $\cc^\atom$ and $\cc^\func$ in the equation below:
\begin{align}
\mathbf{y}(M) = \text{MLP}\Big(\sum_{i \,\in\, \V^\atom} \sigma(\h^\atom_i) \, \phi{(\cc^\atom_i)}, \;  \sum_{u \,\in\, \V^\func} \sigma(\h^\func_u) \, \phi{(\cc^\func_u)} \Big).
\end{align}
Here, we use a graph pooling operation with a weighted sum to generate chemical embeddings for the same molecule at both the atom and motif levels, represented by the two expressions enclosed in parentheses. The weights, $\sigma(\cdot)$ and $\phi(\cdot)$, which are the sigmoid and hyperbolic tangent functions respectively, act as gating mechanisms. These functions evaluate the significance of a local structure for the molecular properties of interest. Additionally, we incorporate an MLP structure to introduce more nonlinearity and enhance the model's expressivity for regression. This improves the model's ability to capture complex relationships and predict molecular properties more accurately.

While our model is inspired by the group contribution theory, it also goes beyond it. First, instead of disregarding the interconnectivity of local structures, it accounts for the influence of neighboring environments on individual atoms and functional groups, by taking their graph embeddings as input. This is particularly important for properties such as the partial charge of an atom and ionization of a functional group, which vary significantly with the local environment. Second, our model is more expressive. The contribution analysis of local structures by Eq.~(\ref{eq:contribution}-\ref{eq:attention}) does not rely on a simple quadratic expansion. Instead, $\que(\cdot)$, $\key(\cdot)$, $\mathbf{\hat{c}}(\cdot)$ and $\mathbf{\Tilde{c}}(\cdot)$ are all modeled by neural networks, providing the capability to represent contributions up to any form of two-body interactions. Third, the self-attention mechanism in our design allows for non-reciprocity in the interaction, meaning that $a_{ij} \neq a_{ji}$. This circumvents the need to construct a nonreciprocal interaction energy $U_{ij} \neq U_{ji}$, which is often used in models such as the UNIFAC method, but is difficult to justify from the perspective of physical interactions.

To evaluate the performance of our model, we first conduct a validation using the standard molecular dataset QM9~\cite{ramakrishnan2014quantum}, which contains approximately 130k molecules labeled with both quantum and chemical properties. Our model exhibits outstanding data efficiency: when trained on only 5\% (6k samples) of the dataset, our model achieves over 97\% accuracy in predicting the highest occupied molecular orbital (HOMO) and lowest unoccupied molecular orbital (LUMO) energies, as shown in Supplementary Fig. S2. The performance with 6k samples is equal to or better than the best baseline's performance with 110k samples~\cite{faber2017machine,gilmer2017neural}. In addition, it can also accurately predict a variety of thermodynamic properties such as free energy, enthalpy, and heat capacity.

To test our model in domain-specific applications, we continue the study of monomers for adhesive polymeric materials. A small dataset of 600 monomer species is prepared, with relevant properties including cohesive energy $E_\text{coh}$, heat of vaporization $\Delta H_\text{vap}$, isothermal compressibility $\beta$, bulk density $\rho$, radius of gyration $R_g$ and glass transition temperature $T_g$ determined using all-atom molecular dynamics simulations. We train our model on 450 randomly selected monomers and test it on the remaining ones. As illustrated in Fig.~\ref{fig:chem}B, most of the properties except $T_g$ can be predicted with over 92\% accuracy. While using the same dataset as the previous work~\cite{schneider2022silico}, which employed a random forest regression model, our model in this work achieves significantly higher accuracy. Note that, measuring the glass transition temperature in simulations is a challenging task. It involves systematically scanning the temperature and requires the system to reach thermal equilibrium at each point. However, near the transition temperature, the dynamics of the system slow down significantly, causing considerable uncertainty in the measurement. As shown in Supplementary Fig. S3, the measurement uncertainty is comparable to the deviation of our model predictions, suggesting that the performance in predicting $T_g$ is likely limited by the quality of the training data.

\vspace{5mm}
\textbf{Automated pipeline for molecular design}
\smallskip

\begin{figure*}
\includegraphics[width=0.98\textwidth]{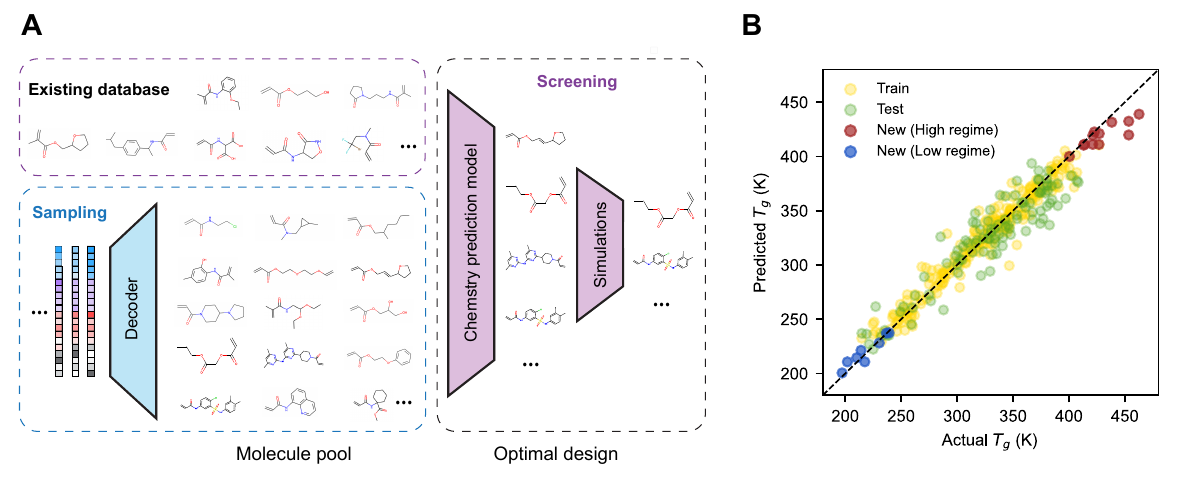}\caption{\textbf{Automated machine learning pipeline for domain-specific molecular design.} \textbf{(A)} We integrate a hierarchical graph autoencoder with a self-attention-based chemistry prediction model to form an autonomous pipeline. The autoencoder employs a coarse-grained functional-group vocabulary to generate chemically valid and structurally diverse candidates, while the chemistry model screens these new molecules based on predicted properties. This division of roles reduces data requirements and increases design flexibility, enabling the pipeline to selectively propose novel compounds aligned with target performance criteria. \textbf{(B)} As a proof of concept, we apply this pipeline to optimize the glass transition temperature (\(T_g\)). The successful identification of new molecules with \(T_g\) values beyond the limits of the training data demonstrates the pipeline’s potential for extrapolation and highlights its effectiveness in guiding data-efficient molecular discovery. } \label{fig:inv}
\end{figure*}

Our chemistry prediction model can process up to $10^4$ molecules in about one hour, yielding property estimates that closely match those from actual simulations. This speed and accuracy facilitate broad and cost-effective high-throughput screening across extensive molecular databases. By employing the model as an initial filter to pinpoint promising candidates, we can then rely on MD simulations to further refine these selections (Fig.~\ref{fig:inv}A). Here, as a case study, we demonstrate the model’s generative power by discovering new polymers with target glass transition temperature (\(T_g\))-- a cornerstone in polymer physics that governs whether a polymer behaves as a rigid solid or a flexible material. Despite its importance for mechanical performance and processing, \(T_g\) remains notoriously difficult to predict due to the intricate interplay of molecular interactions, chain conformation, and free volume; designing polymers with a prescribed \(T_g\) is even more challenging, as there is no simple structure–property rule. Our pipeline addresses this gap by learning the underlying molecular chemistry, achieving high predictive accuracy, and enabling targeted molecular design.

Central to our strategy is a latent-space generative model, which explores new polymer architectures beyond the training database. To expand the molecular search space, we sample molecular embeddings from the prior distribution \(P(\hmol) = \mathcal{N}(\mathbf{0}, \mathbf{I})\). Concretely, we draw random vectors from this standard normal distribution and feed them into the decoder, which converts these latent representations into valid molecular structures biased toward desired property ranges. To highlight the scope of this generative capability, in this fully automated pipeline (Fig.~\ref{fig:inv}A), we constrained the chemical building blocks to acrylate functionalities and sampled 50,000 unique candidates not present in the original database —well above simple enumeration of known compounds. Instead of a brute-force approach, the latent-space model systematically navigates chemical space in a manner that favors valid, non-duplicative structures and spans a broad range, rather than merely generating random permutations.

We next applied our chemistry-prediction model to all 50k generated acrylates, revealing that the predicted \(T_g\) values spanned and extended beyond the range of the training set. We underscore that the model does not rely on random “lucky” hits; rather, it learns molecular motifs associated with low and high \(T_g\). We then screened for materials with particularly high or low \(T_g\) and selected 20 representative candidates for validation via MD simulations. As shown in Fig.~\ref{fig:inv}B, the predicted \(T_g\) values for these candidates are in close agreement with the simulated results, demonstrating our model’s accuracy. Moreover, since the predicted new molecules have \(T_g\) values that exceed the low and high boundaries of the training set, it showcases the model’s capacity to extrapolate beyond known molecules. Although we selected 20 extreme candidates for MD validation, the remaining candidates also followed coherent structural–property trends when examined in aggregate.  Notably, the resulting designs exhibit diverse molecular backbones and functional groups that would be difficult to conceive based on chemical intuition alone.

We recognize that focusing on a single property may appear narrow relative to the size of the sampled set. However, we selected \(T_g\) as a pivotal proof-of-concept target: its sensitivity to both local chemistry and long-range chain dynamics poses a stringent test for any materials-design model. Beyond this demonstration, our framework readily generalizes to multi-property optimization, enabling the concurrent pursuit of mechanical strength, thermal stability, ease of synthesis, and additional constraints vital to industrial practice. Such an expanded search will better leverage the capacity to sample large chemical spaces, ensuring that the final designs balance performance with synthetic viability. By systematically blending generative exploration, high-throughput property prediction, and targeted simulation validation, this foundation provides a robust strategy for machine-guided molecular design and paves the way for accelerated innovation in materials design.

\vspace{10mm}

\textbf{Discussion}
\smallskip

Our approach to molecular design builds upon the emerging paradigm of digitizing chemical space and then steering molecular generation through property-prediction models. Yet, it diverges from conventional strategies that rely on a single embedding vector to represent an entire molecule in an end-to-end framework. Instead, we adopt a hierarchical scheme in which local atomic details, coarse-grained functional groups, and global molecular embeddings each play distinct roles. This architecture not only mitigates the information loss often associated with bottleneck autoencoders, but also ensures that relevant chemical features are extracted and leveraged when needed—akin to the multi-scale design principles used in UNet-like image-processing models~\cite{ronneberger2015u}.

In our pipeline, the decoder incrementally assembles new molecules by predicting structural motifs and the specific bonds connecting them, guided by both local embeddings (atoms and functional groups) and a global molecular embedding that orchestrates the overall design. The subsequent screening employs a self-attention-based chemistry prediction model, which capitalizes on the same hierarchical representations to evaluate molecular properties. By shifting the main burden of chemical interpretation to local embeddings, the framework remains data-efficient: the task of learning chemically meaningful representations at the atomic or group level is considerably more tractable than requiring a single global embedding to capture every subtlety of an entire molecule. This design choice proved critical for achieving high accuracy with limited training data.

A key innovation underlying this efficiency is our use of functional groups as a coarse-grained vocabulary for both generation and prediction. Group contribution theory identifies fewer than 100 such groups that recapitulate most relevant chemistries, offering three significant advantages. First, it confines the combinatorial explosion of possible motifs, avoiding large, data-biased dictionaries. Second, it delegates atom-level connectivity to established cheminformatics toolkits such as RDKit~\cite{rdkit}, alleviating the need for the autoencoder to learn this low-level connectivity from scratch. Finally, we embed these functional groups into a self-attention model that unites domain-specific chemical insights with the flexibility of modern neural networks. As a result, the chemistry prediction model achieves near-simulation-level accuracy while being trained on only a few hundred labeled molecules.

Our demonstration of targeted polymer design through this hierarchical framework showcases its potential to guide molecular discovery efficiently, even in sparse-data regimes. Although we focused here on glass transition temperature, the method naturally extends to multi-property optimization—such as mechanical strength, solubility, or synthetic accessibility—by incorporating additional modules or constraints in the network. In this way, the synergy of coarse-grained functional groups and a self-attention-based architecture can be broadly harnessed for designing biomolecules, polymeric materials, and other hybrid chemical systems. We anticipate that the principles detailed in this work will serve as a robust foundation for the community to build ever more sophisticated machine-learning pipelines, illuminating vast expanses of previously unexplored chemical space and driving rapid innovation in materials science.

\vspace{10mm}
\textbf{Acknowledgments}
\smallskip

This work is supported by the National Science Foundation, through grant NSF-DMR-2119672. 


%

\end{document}


\title{Supplementary Information for:\\
Attention-Based Functional-Group Coarse-Graining: \\
A Deep Learning Framework for Molecular Prediction and Design}

\author{Ming Han}
\thanks{These authors contributed equally to this work.}
\affiliation{Pritzker School of Molecular Engineering,
  University of Chicago, Chicago, IL, USA}
\affiliation{James Franck Institute,
  University of Chicago, Chicago, IL, USA}

\author{Ge Sun}
\thanks{These authors contributed equally to this work.}
\affiliation{Pritzker School of Molecular Engineering,
  University of Chicago, Chicago, IL, USA}
\affiliation{Department of Chemical and Biomolecular Engineering, Tandon School of Engineering, New York University, Brooklyn, NY, USA}
\affiliation{Department of Computer Science, Courant Institute of Mathematical Sciences, New York University, New York, NY, USA}
\affiliation{Department of Physics, New York University, New York, NY, USA}

\author{Juan J. de Pablo}
\thanks{Corresponding author. Email: jjd8110@nyu.edu}
\affiliation{Pritzker School of Molecular Engineering,
  University of Chicago, Chicago, IL, USA}
\affiliation{Department of Chemical and Biomolecular Engineering, Tandon School of Engineering, New York University, Brooklyn, NY, USA}
\affiliation{Department of Computer Science, Courant Institute of Mathematical Sciences, New York University, New York, NY, USA}
\affiliation{Department of Physics, New York University, New York, NY, USA}
\affiliation{Materials Science Division, Argonne National Laboratory, Lemont, IL, USA}

\maketitle
\section{Methods}
A molecule is naturally a graph, where atoms are the nodes interconnected via chemical bonds. Compared to Simplified Molecular-Input Line-Entry System (SMILES)~\cite{weininger1988smiles} strings or its variants, such atom graph provides a better representation of molecular topology. Furthermore, by recognizing the functional groups that are included in a molecule, it becomes possible to also represent it at a coarser level by a motif graph.

In the encoding phase, we first encode the local environment of individual atoms in a molecule by applying graph neural networks to the atom graph. The embedding vectors of atoms are then passed into the corresponding functional groups in the motif graph via a multi-layer perception. Subsequently, we can then encode the local environment of individual functional groups by applying graph neural networks to the motif graph. This process can be modeled through Eqs.~(3-6) in the main text. 
Lastly, molecular embedding $\hmol$ is derived based upon the embedding of the root motif $\hf_0$:
\begin{equation}
   \hmol = \boldsymbol{\mu}(\hf_0) + \boldsymbol{\sigma}(\hf_0) \circ \boldsymbol{\epsilon}, \quad \boldsymbol{\epsilon} \sim \mathcal{N}(\mathbf{0}, \mathbf{I})
\end{equation}
Instead of a deterministic mapping, we perform a statistical inference of $\hmol$: we apply MLPs to derive the mean $\boldsymbol{\mu}$ and the standard deviation $\boldsymbol{\sigma}$, and sample the actual value of $\hmol$ via the reparameterizing trick from a white noise $\boldsymbol{\epsilon}$. 
This variational design not only ensures the continuity of the latent space, but also yields an estimate on the posterior distribution $P(\hmol|M) \approx Q(\hmol|M) = \mathcal{N}(\boldsymbol{\mu}, \boldsymbol{\sigma})$.

In the decoding phase, we reconstruct the molecule by iteratively generating a new motif and connecting it to the partial molecular structure built so far. 
Assuming that molecular embedding carries the key information of molecular structure, we progressively modeled this process conditional on the molecular embedding $\hmol$, until completion using Eqs.~(\ref{eq:prob_mol}). The decoding process can be achieved through Eqs.~(7-10) in the main text. 

\begin{equation}
\begin{aligned}
P(M) &= \int \diff \hmol \,P(\hmol)  \, P \big( M \, \bigr| \, \hmol \big) \\
&= \int \diff \hmol \,P(\hmol) \prod_{u} P \Big( M_{\leq u} \, \Bigr| \, M_{\leq u-1}, \, \hmol \Big). \label{eq:prob_mol}
\end{aligned}
\end{equation}

\section{Hierarchical graph representation}
\subsection{Molecular embeddings}

\begin{figure*}
    \centering
    \includegraphics[width=0.7\textwidth]{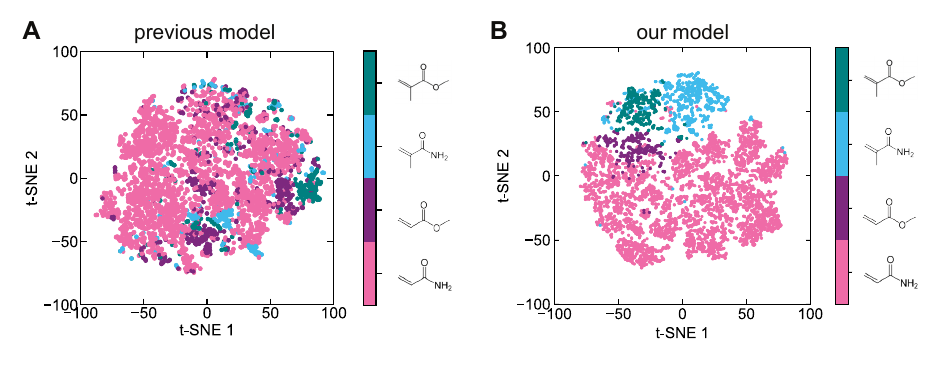} 
    \caption{\textbf{Comparison of model embeddings.} \textbf{(A)} Ineffective clustering in previous model embeddings. The approach relies solely on geometric data extraction of motifs, resulting in reduced chemical relevance and a lack of clear separations among monomer types. \textbf{(B)} Chemically meaningful embeddings in our autoencoder, identical to Fig. 1C in the main text for reference.  We use t-SNE to map ten-dimensional embeddings into a two-dimensional space for both plots.} 
    \label{fig:SI_tsne}
\end{figure*}

Our Hierarchical Graph Autoencoder demonstrates enhanced performance by using a functional-group-based motif vocabulary to create chemically meaningful embeddings. To validate this, we tested the same dataset of 6,000 adhesive monomers with our model and a previous model. The previous model relies on geometrically derived motifs without incorporating prior domain knowledge, resulting in embeddings that lack chemical significance. Specifically, we use t-distributed stochastic neighbor embedding (t-SNE) to map high-dimensional latent space into a two-dimensional space for better visualization. The mapped embeddings fail to form distinct clusters and cannot distinguish different monomer types, as illustrated in Fig.~\ref{fig:SI_tsne}A. To compare, Fig.~\ref{fig:SI_tsne}B shows that our autoencoder effectively groups similar chemical structures into distinct clusters.

\subsection{Reconstruction of molecules}

To evaluate the efficacy of our hierarchical graph autoencoder, we compared its performance against established baselines: a string-based autoencoder~\cite{gomez2018automatic}, an atom-graph-based~\cite{liu2018constrained} autoencoder, and a data-driven motif extraction hierarchical variational autoencoder~\cite{jin2020hierarchical} (HierVAE). As shown in Table~\ref{tab:methods_comparison}, our model reconstructs molecular structures with a success rate of approximately 95\%, while the first two baseline methods struggled to exceed 50\% under comparable conditions. Moreover, our method outperforms HierVAE by leveraging fewer than 100 well-established functional groups, which serve as a chemically meaningful vocabulary that faithfully captures the reactivity and structural diversity of polymeric and specialty chemicals. In contrast, purely data-driven motif extraction strategies often bias the model toward frequent ring motifs, fragmenting moiety into single/double bonds, thereby underserving chain-like and branched topologies and reducing overall fidelity.

Furthermore, our representation systematically integrates chemical information at multiple scales: (i) local atomic embeddings that capture fine-grained structural details, (ii) coarse-grained functional groups that encode chemically relevant building blocks, and (iii) a global molecular embedding that situates each molecule within the broader design space. This multi-level structure not only ensures robust reconstructions but also yields chemically interpretable latent features (see Fig. 1C), a marked improvement over existing data-driven hierarchical autoencoder models.

\begin{table}[h!]
    \centering
    \caption{Comparison of methods}
    \label{tab:methods_comparison}
    \begin{tabular}{ccc}
        \toprule
        \multirow{2}{*}{Method} & \multicolumn{2}{c}{Reconstruction}  \\
        \cmidrule(lr){2-3} 
        & Accuracy & Valid  \\
        \midrule
        Actual data & - & 100\%  \\
        SMILES VAE & 29.7\% & 94.5\%  \\
        Graph-based VAE & 48.2\% & 100\%  \\
        HierVAE & 76.9\% & 100\%  \\
        \midrule 
        Our work & \textbf{95.3\%} & 100\%  \\
        \bottomrule
    \end{tabular}
\end{table}

\section{Chemistry Prediction performance}
\subsection{Benchmark: QM9 dataset}
To investigate the success of our model in predicting chemical properties, we used the publicly available QM9 dataset~\cite{ramakrishnan2014quantum}. The QM9 dataset is a widely used benchmark in computational chemistry and machine learning for quantum chemistry, consisting of approximately 130,000 molecules with detailed quantum chemical properties. It provides a comprehensive platform for evaluating the predictive capabilities of machine learning models. Previous studies using the QM9 dataset have employed various machine learning techniques, such as regressors~\cite{faber2017machine}, graph neural networks~\cite{faber2017machine,rahaman2020deep}, and message-passing neural networks~\cite{gilmer2017neural}, to predict chemical and thermodynamic properties. However, these models typically require a large portion of the dataset for training to achieve high accuracy~\cite{faber2017machine}.

In contrast, our model demonstrates exceptional data efficiency. From the dataset, we randomly select 6000 molecules for training and another 2000 for testing. As shown in Supplementary Fig.~\ref{fig:SI_qm9}, even when trained on only 5\% of the QM9 dataset, our model achieves over 97\% accuracy in predicting the energies of the HOMO and LUMO. The ability of our model to perform so well with limited training data can be attributed to its model architecture, which effectively captures the underlying chemical information of the molecules. This contrasts with traditional methods that rely heavily on large datasets to achieve similar levels of accuracy. The high accuracy achieved by our model with minimal data usage not only demonstrates its robustness but also suggests potential real-world applications in scenarios where data availability is limited.

\begin{figure*}
    \centering
    \includegraphics[width=0.6\textwidth]{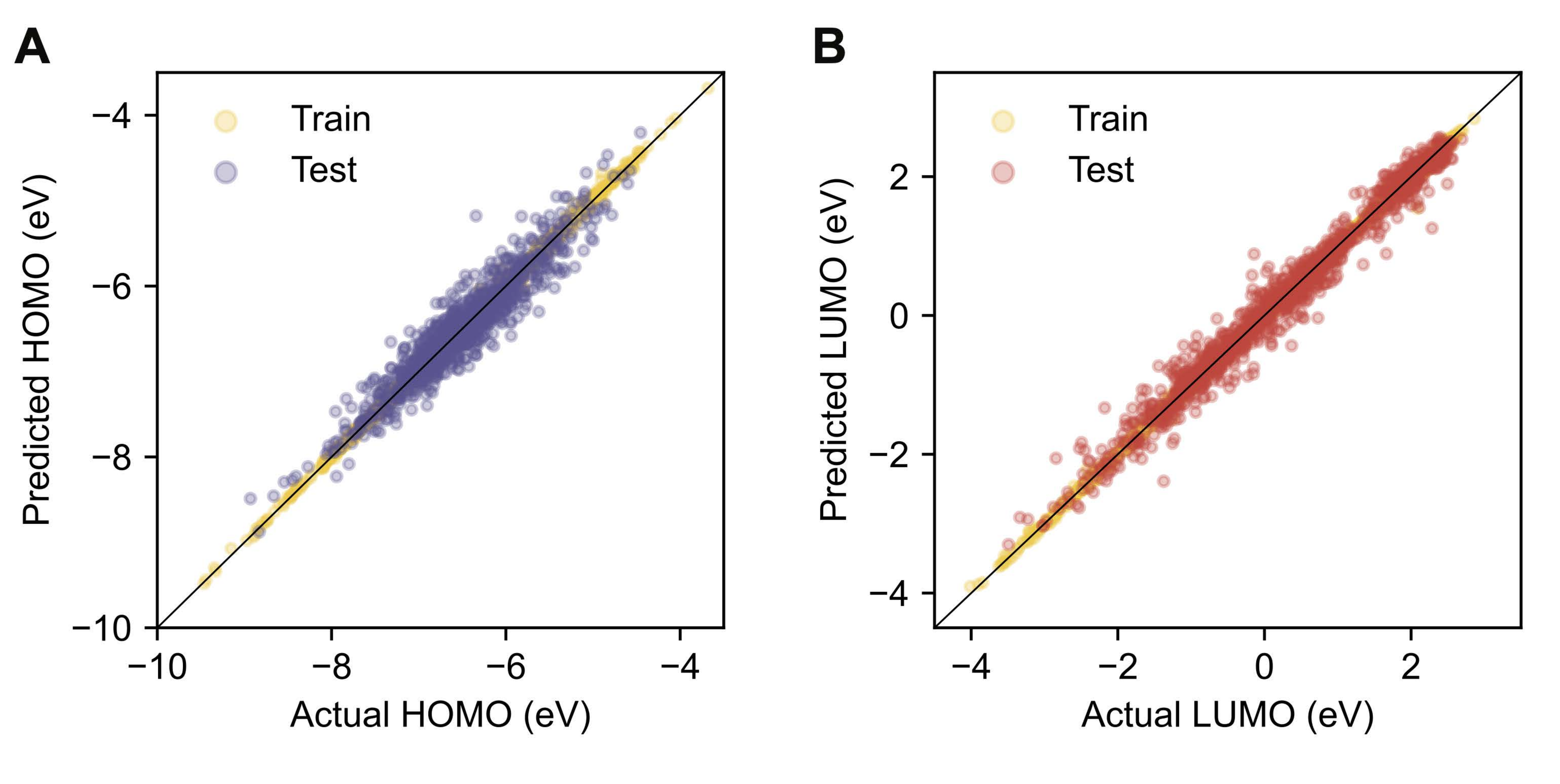} 
    \caption{\textbf{Prediction accuracy of the model on the QM9 dataset.} \textbf{(A)} Highest occupied molecular orbital (HOMO) energy predictions and \textbf{(B)} lowest unoccupied molecular orbital (LUMO) energy predictions. The model achieves 97\% accuracy in both cases, despite being trained on only 5\% of the dataset. This demonstrates the robust performance of the model with limited training data.} 
    \label{fig:SI_qm9}
\end{figure*}

\subsection{Application: Monomers for adhesive polymeric materials}
The accurate prediction of thermodynamic properties is critical for designing polymeric materials with desired characteristics. However, some datasets are challenging to prepare. For example, measuring the glass transition temperature ($T_g$) in simulations requires systematic temperature scanning and achieving thermal equilibrium at each point. Near the transition temperature, the system's dynamics slow down significantly, introducing considerable uncertainty to the measurements.

\begin{figure*}
    \centering
    \includegraphics[width=0.4\textwidth]{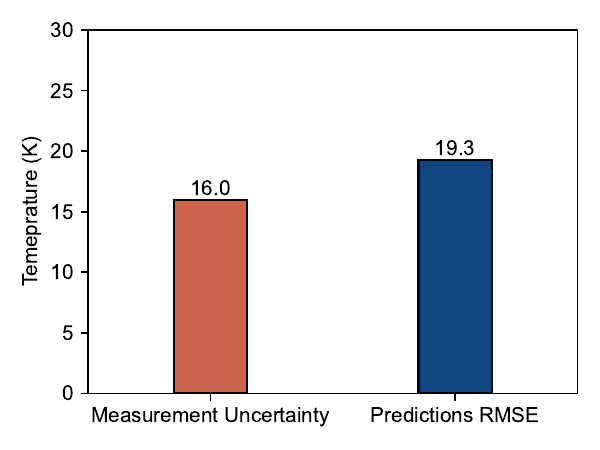} 
    \caption{\textbf{Measurement uncertainty and prediction RMSE for glass transition temperature ($T_g$).} The measurement uncertainty and root mean square error (RMSE) of our model predictions for $T_g$ are comparable. This indicates that the performance is likely limited by the quality of the training data, which is inherently challenging to generate through simulations.} 
    \label{fig:SI_tg}
\end{figure*}

In Fig. 2B in the main text, we demonstrate that our model can accurately predict cohesive energy, heat of vaporization, isothermal compressibility, bulk density, and radius of gyration. However, although still significant, the prediction accuracy for $T_g$ does not reach the same high levels as those for the other properties. Supplementary Fig.~\ref{fig:SI_tg} compares the measurement uncertainty with the RMSE of our model's predictions for $T_g$. The comparable values indicate that the primary limitation of our model's predictive performance is the quality of the training data, rather than deficiencies in the model itself. This underscores the inherent difficulty in generating accurate $T_g$ data through simulations. Despite this, our model performs very well for other properties, highlighting its robustness and effectiveness.

\clearpage
\section*{Supplementary References}
%